\documentclass{emulateapj}
\begin{document}

\slugcomment{ApJ Letters, accepted}

\title{Ultra-Compact Dwarfs in the Fossil Group NGC 1132\altaffilmark{1}}

%---------------------------------------------------------------------

\author{Juan P. Madrid\altaffilmark{2}}

\altaffiltext{1}{Based on observations made with the NASA/ESA Hubble
Space Telescope, obtained at the Space Telescope Science Institute,
which is operated by AURA, Inc., under NASA contract NAS5-26555.  
These observations are associated with program 10558.}

\altaffiltext{2}{Centre for Astrophysics and Supercomputing, Swinburne
University of Technology, Hawthorn, VIC 3122, Australia}

%-----------------------------------------------------------------------

%-----------------------------------------------------------------------

\begin{abstract}

Eleven Ultra-Compact Dwarf (UCD) and 39 extended star cluster candidates are found to 
be associated with the galaxy NGC 1132. This giant elliptical galaxy is the remnant of a 
fossil group. UCD and extended star cluster candidates are  identified through the analysis 
of their structural parameters, colors, spatial distribution, and luminosity using deep 
Hubble Space Telescope observations in two filters: the F475W (Sloan $g$) and F850LP 
(Sloan $z$). The median effective radius of these UCDs is $r_h=13.0$ pc. Two types 
of UCDs are identified in the vicinity of NGC 1132, one type shares the same color 
and luminosity of the brightest globular clusters and traces the onset of the mass-size 
relation. The second kind of UCD is represented by the brightest UCD candidate,  
a M32-type object, with an effective radius of  $r_h=77.1$ pc, located at $\sim 6.6$ kpc 
from the nucleus of NGC 1132. This UCD candidate is likely the remaining nucleus
of a minor merger with the host galaxy. With the exception of a particularly blue UCD candidate,
UCDs are found to extend the mass-metallicity relation found in globular clusters to 
higher luminosities. The results of this work support the growing body of evidence 
showing that UCDs are not circumscribed to galaxy clusters as previously thought. 
UCDs are likely to be a common occurrence in all environments. The milder tidal field
of a fossil group, when compared to a galaxy cluster, allows UCDs and extended star 
clusters to survive up to present time at small galactocentric distances.

\end{abstract}

\keywords{galaxies: star clusters - galaxies: elliptical and
lenticular, cD - galaxies: clusters: general - galaxies: dwarf - 
galaxies: groups: individual (NGC 1132)}

%-----------------------------------------------------------------------
%-----------------------------------------------------------------------
%-----------------------------------------------------------------------
\maketitle

\section{Introduction}

Ultra-Compact Dwarfs are low mass stellar systems with properties in between those
of bright globular clusters and the most compact dwarf elliptical galaxies. Since their 
discovery UCDs have been linked to galaxy clusters. Initially found in the Fornax 
cluster (Hilker et al.\ 1999; Drinkwater et al.\ 2000) UCDs have been reported in 
Virgo, Centaurus, Hydra, Coma and the more distant 
clusters Abell S040 and Abell 1689 (Hasegan et al. 2005; Evstigneeva et al.\ 2008; 
Chilingarian \& Mamon 2008; Mieske et al.\ 2004, 2007; Wehner \& Harris 2007; 
Misgeld et al.\ 2008; Madrid et al.\ 2010; Blakeslee \& Barber DeGraaff 2008). 

The search for UCDs has been focused in galaxy clusters and observations seeking to 
discover UCDs in other environments are critically needed to understand their 
formation process and to determine how common these low mass stellar systems truly are. 
Only recent accounts of UCDs outside galaxy clusters with spectroscopic 
confirmation exist in the literature. Evstigneeva et al.\ (2007) search for UCDs in 
six galaxy groups using ground based data and found only one definite candidate in 
the Dorado group. Romanowsky et al.\ (2009) describe UCDs in the galaxy group of 
NGC 1407 while Hau et al.\ (2009) find a UCD located in the vicinity of the 
Sombrero Galaxy, a low density environment. Da Rocha et al.\ (2011) give a detailed 
account of the UCDs present in two Hickson compact groups. Recently, and after discovering 
UCDs in field (NGC 4546) and group (NGC 3923) environments Norris \& Kannappan (2011) 
claim the ubiquity of UCDs in galaxies with populous globular cluster systems (GCS).

A search for UCDs is undertaken in the isolated giant elliptical NGC 1132.
This galaxy is likely the only remnant of a merged galaxy group, or fossil group.
Fossil groups have a defining characteristic gap of two magnitudes between the brightest
galaxy and the second brightest galaxy. The temperature,  metallicity, and luminosity 
of the X-ray halo surrounding NGC 1132 are similar to those of galaxy groups 
(Mulchaey \& Zabludoff 1999). Analysis of the Millenium Simulation (Springel et al.\ 2005) 
show that fossil groups assembled their dark matter halos early, accumulating 80\% of the 
present day mass 4 Gyr ago (Dariush et al.\ 2007). The absence of parent galaxies and thus 
lack of substructure in fossil groups has been debated as a possible challenge to the 
$\Lambda$CDM paradigm, similar to the missing galaxy problem of the Local Group 
(Sales et al.\ 2007 and references therein). 

Fossil groups are dynamically evolved environments and due to their early assembly 
and their fast evolution the dominant galaxy has had time to accrete the most massive 
surrounding galaxies through dynamical friction (Mulchaey \& Zabludoff 1999). 
Dwarf galaxies are  exempt from the consequences of dynamical friction given that 
this effect is proportional to the mass of the satellite galaxy. Also, given that 
ram pressure stripping is milder in a fossil group than in the core of a galaxy cluster, 
UCDs formed by downsizing dwarf galaxies (Bekki et al.\ 2003) are expected to survive up to 
the present day in fossil groups.

This work uses the same method employed by Madrid et al.\ 2010 (hereafter Paper I)
to determine the presence of UCDs in the Coma Cluster. The brightest UCD candidates 
reported in Paper I have been spectroscopically confirmed by Chiboucas et al.\ (2010) 
using the Keck telescope. The fainter end of the luminosity distribution of UCD candidates 
presented in Paper I lacks spectroscopy confirmation simply due to the onerous challenge 
of obtaining their spectra at the Coma distance. 

Collobert et al.\ (2006) give a redshift of $z=0.023$ for NGC 1132, 
and distance modulus of (m-M)=34.86 mag  equivalent to a luminosity distance 
of 99.5 Mpc using H$_0$=71 km.s$^{-1}$Mpc$^{-1}$, $\Omega_M=0.27$, 
and $\Omega_\Lambda=0.73$ (Wright 2006). 

%-----------------------------------------------------------------------
%-----------------------------------------------------------------------

\section{Data and reductions} % Section 1

The observations of NGC 1132 are carried out with the Advanced Camera for Surveys
Wide Field Channel (ACS/WFC) on 2006 August 22. The data is acquired in two bands: 
$F475W$, similar to Sloan $g$, and $F850LP$, Sloan $z$ (Mack et al.\ 2003). Four exposures are
obtained in each band for a total exposure time of 4446 $s$ in $F475W$ and  
6885 $s$ in $F850LP$. The pixel scale of the ACS/WFC is $0.\arcsec05$/pixel (23pc/pixel) and the
physical scale of the field of view is $\sim93\times93$ kpc at the distance of
NGC 1132. The same physical scale is covered in the study of Paper I.

Eight flatfielded science images ({\sc flt.fits}) are retrieved from the public
archive hosted at the Space Telescope Science Institute. These files are 
pre-processed  through the standard pipeline that corrects bias, dark current, 
and flatfield (Sirianni et al.\ 2005). The {\sc pyraf} task 
{\sc multidrizzle} is used to combine the different exposures within 
the same filter, remove cosmic rays, and perform distortion correction.

Detection of sources is carried out with SExtractor (Bertin \& Arnouts 1996)
using a detection threshold of 3$\sigma$ above the background. A careful round of visual 
inspection allows for the discarding of contaminants such as background galaxies. 
As recently shown by the Galaxy Zoo project the human eye remains a powerful tool 
to discriminate between different types of galaxy morphology (Lintott et al.\ 2011).

As in Paper I the empirical point spread function (PSF) is created by running 
the {\sc pyraf} tasks {\sc pstselect}, {\sc psf}, and {\sc seepsf}. These tasks 
respectively select bright unsaturated stars from the image, build the PSF by fitting
an analytical model (Moffat function), and subsample the PSF as required by 
the software used to determine the structural parameters.

%---------------------------------------------------------------------

\section{Analysis}

%---------------------------------------------------------------------

\subsection{Structural Parameters}

Structural parameters are determined using {\sc ISHAPE} (Larsen 1999). This 
specialized software convolves the PSF with an analytical model of the surface
brightness profile of slightly resolved stellar structures such as globular 
clusters and yields their effective radius, ellipticity, position angle, 
signal-to-noise ratio, and an estimate of the goodness of fit.
An important parameter to consider when studying low mass stellar systems 
in different environments is the effective radius or half-light radius ($r_h$). 
A King profile (King 1962, 1966) with a concentration parameter, 
or tidal to core radius, of c=30 is the most suitable analytical model for fitting 
the surface brightness profile of globular clusters and UCDs (see Paper I).
The structural parameters of 1649 point-like sources are derived in both bands. 
The values of structural parameters quoted hereafter are those corresponding
to the $F850LP$ band given its longer exposure time. This filter is also a better tracer 
of the mass of globular clusters and UCDs.

%---------------------------------------------------------------------

\subsection{Photometry}

Photometry is performed using an aperture of 4 pixel in radius with the task {\sc phot} 
within the {\sc daophot} package of {\sc pyraf}, aperture correction is applied using 
the formulas of Sirianni et al. (2005). Photometric measurements are obtained 
for all 1649 stellar objects identified as globular clusters and UCD candidates
in the ACS images of NGC 1132. 

Foreground extinction for NGC 1132 is $E(B-V)=0.063$ mag (from NED), the 
specific extinction corrections applied to each HST filter are obtained following 
the prescriptions of Sirianni et al.\ (2005) which give $A_{F475W}=0.229$ mag and 
$A_{F850LP}=0.094$ mag. Up-to-date photometric zeropoints for ACS/WFC are 
obtained from the STScI website: $F475W=26.163$, $F850LP=24.323$. 
These zeropoints yield magnitudes in the Vega magnitude system.

%---------------------------------------------------------------------

\subsection{Selection Criteria}

UCDs are selected based on their color, magnitude and structural parameters.
A metallicity break found by Mieske et al.\ (2006) at the onset of the luminosity-size 
relation sets at $M_V=-11$ the luminosity cutoff to distinguish between simply 
bright globular clusters and UCDs. At the distance of NGC 1132 
this luminosity requirement translates to $m_{F850LP} <$ 22.3 mag, this value 
is obtained using {\sc calcphot} within the {\sc synphot} package of 
{\sc stsdas} (Laidler et al.\ 2005). In color, UCD candidates must be in the 
same range defined by the globular cluster system, this is shown in the next 
section to be $1< (F475W-F850LP) < 2.5$

In size, UCDs are conventionally, but arbitrarily, defined as stellar 
systems having an effective radius between 10 and 100 parsecs (see Paper I 
and references therein). UCDs must be consistently resolved in both bands, 
at the distance of 100 Mpc  this condition is satisfied for sources with 
$r_h > 8$pc with the deep images analyzed in this work.

In photometric studies, and as a safeguard against contamination by 
background galaxies, the ellipticity of UCDs is expected to be between
$\epsilon = 0$ and $\epsilon = 0.5$ (Blakeslee \& Barber de Graff 2008, 
Paper I). All UCD candidates fall within that range of ellipticity.
Additionally, a signal-to-noise ratio of $S/N >50$ is a requirement for 
obtaining a reliable estimate of structural parameters with {\sc ISHAPE} 
(Harris 2009).

%---------------------------------------------------------------------

\begin{figure}

\plotone{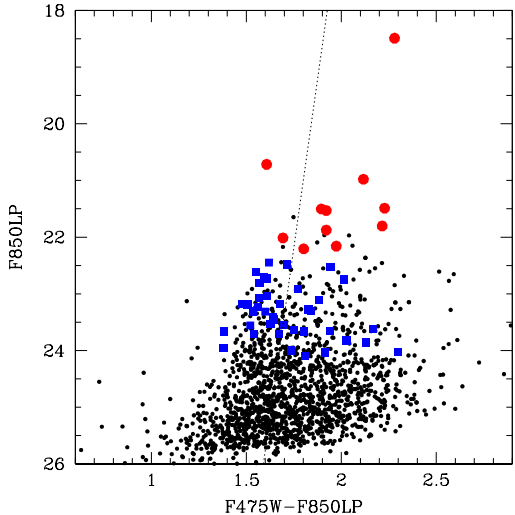}

\caption{Color-Magnitude Diagram of the globular cluster system associated 
with the isolated elliptical NGC 1132. Unresolved globular clusters are plotted as 
black dots. UCD candidates are represented by red circles and extended clusters 
are plotted as blue squares. The dashed line shows the Mass Metallicity
Relation with a slope of $\gamma_z=0.042$, see Peng et al.\ (2009) for a discussion. 
\label{fig2}}
\end{figure}

%---------------------------------------------------------------------
%---------------------------------------------------------------------

\section{Results}
%---------------------------------------------------------------------
%---------------------------------------------------------------------

The Color-Magnitude Diagram (CMD) of the globular cluster system 
associated with NGC 1132 is used to define the parameter space 
where UCDs are expected to be present, a detailed analysis of this CMD is 
not the aim of this paper. For most globular clusters 
($99\%$ of them) their colors are in the range $1<(F475W-F850LP)<2.5$ and their 
magnitudes between $21<m_{F850LP}<26$ as shown in Figure 1. This broad color 
range translates into a metallicity range of $-2.61<[Fe/H]<-0.10$ given 
by the color-metallicity transformation of Peng et al.\ (2006). The deduced 
metallicity range is in agreement with previous metallicity 
estimates of UCDs (Mieske et al.\ 2004, Evstigneeva et al.\ 2007).

%---------------------------------------------------------------------
%---------------------------------------------------------------------

The vast majority of 1649 globular clusters and UCD candidates identified in the ACS images
remain unresolved or their $S/N$ is too low to obtain a truly reliable estimate of their structural 
parameters. However, 11 sources satisfy the color, magnitude, size, 
and $S/N$ criteria outlined above to be classified as UCD candidates. The effective 
radii of these 11 UCD candidates range from 77.1 pc, for the largest and brightest candidate, 
to 8.5 pc. The median effective radius for UCDs is $r_h =13.0$ pc with a standard deviation 
of $\sigma =19.8$ pc. Excluding the brightest candidate brings these numbers to $r_h =11.4$ pc 
and $\sigma =5.1$ pc. All 11 UCD candidates have a $S/N>90$ and are plotted as red dots 
in the color-magnitude diagram of Figure 1. In color and magnitude most UCDs overlap with the 
brightest globular clusters, the very same is the case for UCDs in the core of the Coma cluster, 
i.\ e.\ they share the same parameter space of the brightest globular clusters associated 
with NGC 4874 (Paper I). 

%---------------------------------------------------------------------
%---------------------------------------------------------------------

%---------------------------------------------------------------------

\begin{figure}
\plotone{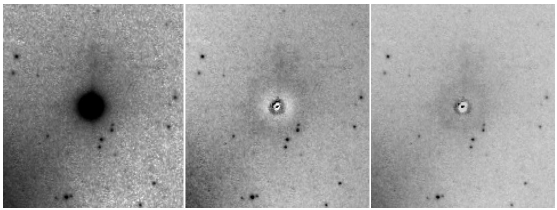}

\caption{M32 counterpart in the Fossil Group NGC 1132. From left to right: original 
$F850LP$ image; residual of best fit using {\sc galfit} and a Sersic model with 
$n$=2.4; residual of best fit using {\sc ishape} and a King model with $c=30$.
\label{fig4}}
\end{figure}

%---------------------------------------------------------------------
\subsection{A M32 equivalent}

The brightest UCD candidate is a M32-like object located, in projection, 
$\sim$6.6 kpc away from the center of the galaxy. Its effective radius is $r_h=77.1$ pc, 
its magnitude $m_{F850LP}=18.49$, or two magnitudes brighter than the second brightest 
UCD candidate, and its is color particularly red $(F475W-F850LP)= 2.28$. In the framework 
of a dual formation mechanism for UCDs proposed by Da Rocha et al.\ (2011) 
and Norris \& Kannappan (2011) this candidate is the remaining nucleus of a 
stripped companion of NGC 1132 due to the clear gap of two magnitudes between
the brightest globular clusters and this UCD. This object is similar to SDSS J124155.3+114003.7,
the UCD reported by Chilingarian \& Mamon (2008) at a distance of 9 kpc from M59.

As stated by Norris \& Kannappan (2011) no globular cluster system has a luminosity 
function continuously extending up to such high luminosity. As shown in Paper I, 
not even the extremely rich globular cluster system of NGC 4874 in the core of the 
Coma cluster has members with magnitudes similar to the brightest UCD candidate.
This object was catalogued by the Two Micron All Sky Survey as 2MASS-02525121-0116193,
its K band magnitude is $M_K=13.505$ mag  (Skrutskie et al.\ 2006). 

A size estimate for the M32 counterpart was also carried out with {\sc galfit} (Peng et al.\ 2002)
using a Sersic model with an initial Sersic index of $n=2$ (Sersic 1968).  Using {\sc galfit} 
the best fit for this object has a reduced $\chi^2$=1.008 and yields $n=2.4$, and 
$r_h=89.2$ pc. The original image of the brightest UCD and the residual after model 
subtraction with both {\sc galfit} and {\sc ishape} is given in Figure 2. The best 
fit of two different analytical models and different software leaves a small residual 
in the core. Similar residuals are found by Price et al.\ (2009) studying several 
compact elliptical galaxies in the core of the Coma Cluster. The brightest UCD candidate
is comparable, but slightly smaller than the compact elliptical galaxies studied by 
Price et al. (2009) which have effective radii of $r_h\sim200$ pc. This object is also 
fainter than the Price compact ellipticals which have magnitudes ranging from $m_B=18.34$ to 
$m_B=21.33$ mag. Photometric and structural parameters of the eleven UCD candidates are 
presented in Table 1.
%---------------------------------------------------------------------
%---------------------------------------------------------------------

\subsection{Extended Star Clusters}

An additional 39 sources are positively resolved in both bands and have effective 
radii between 8.2 pc and 59.7 pc, with a median $r_h =11.6$ pc. While these sources
have the size characteristic of UCDs their magnitudes fall short of the minimum
threshold for the selection criteria. These extended stellar systems have received 
several denominations, here we refer to them as extended star clusters.
Extended star clusters are plotted as blue squares on the CMD of Figure 1. 
Even if these extended star clusters are not as bright as UCDs, in luminosity 
some of these objects are comparable to $\omega$ Centauri.
%---------------------------------------------------------------------

%---------------------------------------------------------------------

\begin{figure}
\plotone{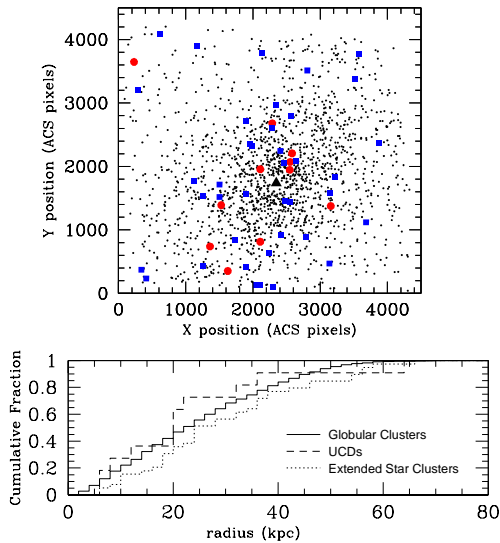}

\caption{Spatial distribution of compact stellar systems in the Fossil Group NGC 1132.
{\it Top panel}: positions of all globular clusters as black dots, extended star clusters 
as blue squares and UCDs as red circles.  Both panels show that globular clusters, extended star 
clusters, and UCDs congregate around the central galaxy designated by a black triangle 
($x=2347, y=1740$) in the center of the top panel. The brightest UCD is located 6.6 kpc 
from the center of the galaxy $x=2547, y=1946$. {\it Bottom panel}: Cumulative fraction for all the 
stellar systems listed above. The maximum difference with a cumulative fraction between the 
different stellar systems is greater than in Coma (Paper I, Fig. 10). \label{fig5}}
\end{figure}

%---------------------------------------------------------------------
%---------------------------------------------------------------------

\section{Spatial Distribution}

The position of globular clusters, extended star clusters, and UCDs on the ACS frame are 
plotted in the top panel of Figure 3. All these low-mass stellar systems congregate towards
the central elliptical arguing against background contamination, this is also 
shown in the bottom panel. Five UCDs and seven extended star clusters are found 
within the inner 12 kpc (in projection) to the center of the galaxy. In Paper I 
no UCDs were found within the inner 15 kpc of the center of NGC 4874. As noted 
in the introduction a milder tidal field and ram pressure stripping within a 
fossil group can allow UCDs and extended star clusters to survive up to the 
present day at small galactocentric distances.
%---------------------------------------------------------------------

\begin{figure}
\plotone{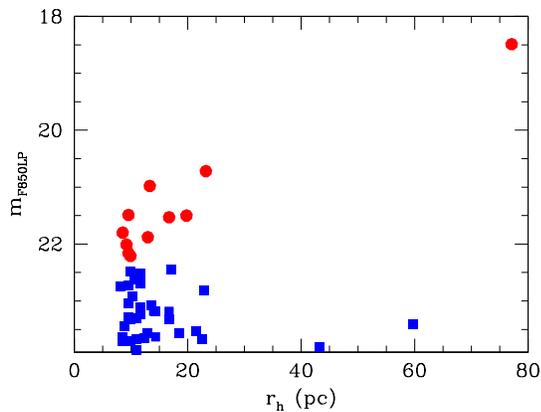}

\caption{Magnitude vs. size of the 11 UCD  and 39 extended star cluster candidates 
associated with NGC 1132. UCDs trace the onset of a correlation between size and 
luminosity absent for extended star clusters and globular clusters. \label{fig3}}
\end{figure}

%---------------------------------------------------------------------

%---------------------------------------------------------------------

\begin{deluxetable}{ccccc}
\tabletypesize{\footnotesize}
\tablecaption{Structural and photometric parameters of 
UCD Candidates \label{tbl-1}} 

\tablewidth{0pt} \tablehead{
\colhead{$r_{h F850LP}$ (pc)} & \colhead{$\epsilon$} & \colhead{S/N} & \colhead{$m_{F850LP}$}  & \colhead{Color}
}
 
\startdata

77.1  &   0.09  & 1168 &   18.49  &   2.28 \\
23.2  &   0.11  &  323 &   20.72  &   1.61 \\
13.3  &   0.22  &  287 &   20.98  &   2.12 \\
9.6   &   0.16  &  164 &   21.49  &   2.23 \\
19.8  &   0.37  &  213 &   21.50  &   1.89 \\
16.7  &   0.05  &  190 &   21.53  &   1.92 \\
8.5   &   0.25  &  166 &   21.80  &   2.22 \\
13.0  &   0.10  &  153 &   21.88  &   1.92\\
9.2   &   0.08  &  133 &   22.01  &   1.69\\
9.6   &   0.12  &   89 &   22.16  &   1.97\\
9.9   &   0.33  &   95 &   22.21  &   1.80\\

 \enddata

 \tablecomments{Column 1: effective radius in pc measured in the F850LP band; 
 Column 2: ellipticity; Column 3: signal-to-noise ratio; Column 4: F850LP magnitude;
 Column 5: F475W-F850LP color.}

\end{deluxetable}

%---------------------------------------------------------------------

%---------------------------------------------------------------------
%---------------------------------------------------------------------

\section{Magnitude-Size relation, Mass-Metallicity relation}

One of the defining characteristics of UCDs is their magnitude-size relation. UCDs
are the smallest stellar systems that show a correlation between luminosity (or mass)
and size (Hasegan et al.\ 2005), with masses  of $\sim 2\times 10^6 M_\odot$ these objects 
are indeed at the onset of this relation. In Figure 4 the magnitude-size relation for the 11
UCD candidates plotted as red cirlces is evident. Magnitude and $r_h$ for the 11 UCDs in 
Figure 4 have a Spearman rank correlation coefficient of $\rho=0.7$. The correlation 
is not perfect (i.\ e.\ $\rho = 1$) due to the scatter characteristic of such plots. 
No magnitude-size correlation is found for extended star clusters, with $\rho=0.15$,
these objects are plotted as blue squares in Figure 4. The effective radius of globular clusters 
with masses below $\sim 2\times 10^6 M_\odot$ do not show any correlation with luminosity either, with 
a coefficient of $\rho =0.1$ derived by McLaughlin (2000), that is, almost completely 
uncorrelated (i.\ e.\ $\rho = 0$). 

A second correlation for stellar systems with masses of  $\sim 10^6 M_\odot$
and above is the mass-metallicity relation (MMR, Harris et al.\ 2006, Strader et al.\
2006). UCDs also obey this relation. The MMR is evident in two simple ways
in the CMD of Figure 1. At the distance of NGC 1132 ($\sim 100 $ Mpc) the paradigmatic bimodality of 
GCS is blurred, however if we consider the measurements presented by Da Rocha et al.\ (2011) 
for GCS of compact groups at lower redshift the color of the metal-poor subpopulation of
globular clusters is $(F475W-F850LP)\sim 1.6$. In the CMD of Figure 1 blue globular clusters
do not extend to luminosities higher than $m_{F850LP} \sim 22.5$. With only one exception,
no clusters or extended  objects are found to have the colors of blue globular clusters and 
luminosities higher than $m_{F850LP} \sim 22.5$. The second brightest UCD, located 20.6 kpc
away from the galaxy center, is exceptionally  
blue (F475W-F850LP)=1.61 and stands out of the MMR.

A second manifestation of the MMR quantified in a simple way is the median color of UCDs and 
extended star clusters. Self-enrichment models (Bailin \& Harris 2009) predict that more massive 
(and bigger) objects migrate in color towards the red. UCDs that are more massive than extended 
globular clusters have indeed redder colors on average. The median color for UCDs is 
$(F475W-F850LP)= 1.92$ with $\sigma=0.22$ while the median color for extended globular clusters 
is $(F475W-F850LP)= 1.67, \sigma=0.21$. The brightest UCD candidate is also the reddest with a 
color of $(F475W-F850LP)= 2.28$, this is in agreement with the findings of 
Norris \& Kannappan (2011). In contrast, the largest extended globular cluster with $r_h=59.7 pc$ 
has a blue color with $(F475W-F850LP)= 1.64$. UCDs extend the MMR to brighter magnitudes 
than the end of the globular cluster luminosity function.\\

%---------------------------------------------------------------------

\section{Discussion}

In Paper I and in this study two highly evolved environments with fundamentally 
different densities are probed: the Core of the Coma Cluster, the richest galaxy 
cluster of the nearby Universe and an isolated elliptical galaxy in a fossil group. Both systems 
exhibit similar characteristics of their UCD population with the exception 
that UCDs are present at small galactocentric distance (projected) in the 
fossil group and not in the galaxy cluster.

The presence of UCDs at the bright tip of the globular cluster systems
of elliptical galaxies appears independent of environment. Given also the 
evidence given by the works cited in Section 1 on the presence of UCDs 
outside galaxy clusters it is natural to believe that UCDs are commonly 
present in all environments. The results of this work support the thesis
put forward by Norris \& Kannappan (2011) on the ubiquity of UCDs in 
galaxies with populous globular cluster systems irrespective of environment.

Only the superior resolution of the Hubble Space Telescope allows for the determination 
of the structural parameters of UCDs in the local Universe.  Evstigneeva et al.\ (2007)
searched for UCDs in five galaxy groups to no avail using ground based data.
The fact that UCDs escaped detection for almost a decade while HST was operational 
is not a coincidence. HST can resolve UCDs only at relatively low redshift 
(D$\sim$ tens to hundreds of Mpc) however at these distances first and second 
generation instruments onboard HST (e.g. WFPC2) have small field of views that only 
cover physical scales of $\sim10\times10$ kpc.  UCDs are difficult to find 
within the innermost 10 kpc of the host galaxy (Paper I, Bekki et al.\ 2003, 
Da Rocha et al.\ 2011) not only due to the crowding produced by starlight but also 
due to the destructive effects of the tidal field of the host galaxy 
(Bekki et al.\ 2003). An artificial maximum luminosity cutoff usually used in 
globular cluster studies has also certainly contributed to under-reporting of 
UCDs in the past (Norris \& Kannappan 2011).

Spectroscopic confirmation of the UCD candidates presented in this work 
is needed and will be actively sought with a 10m telescope. 

%---------------------------------------------------------------------
%---------------------------------------------------------------------
%---------------------------------------------------------------------

%---------------------------------------------------------------------
%---------------------------------------------------------------------

\acknowledgments

Many thanks to L. Spitler, J. Hurley, W. Harris, L. Schwartz, and D. Croton  for their input. 
The referee gave a detailed and constructive report that helped to improve the manuscript. 
This research has made use of the NASA/ADS, the NASA/IPAC Extragalactic Database, SIMBAD, 
and Google. STSDAS and PyRAF are products of the STScI, which is operated 
by AURA for NASA.

%---------------------------------------------------------------------

{\it Facilities:} \facility{HST (ACS)}

%---------------------------------------------------------------------


\begin{thebibliography}{}

\bibitem[Bailin(2009)]{bai09} Bailin, J. \& Harris, W. E. 2009, ApJ, 695, 1082

\bibitem[Bertin(1996)]{ber96} Bertin, E., \& Arnouts, S. 1996, A\&AS, 117, 393

\bibitem[Blakeslee \& Barber de Graaff(2008)]{bla08} Blakeslee,
J. P. \& Barber de Graaff, R. 2008, AJ, 136, 2295


\bibitem[Bekki et al.(2003)]{bek03} Bekki, K., Couch, W. J., Drinkwater, M. J., \&  Shioya, Y. 2003, MNRAS, 344, 399

\bibitem[Collobert et al.(2006)]{col06} Collobert, M., Sarzi, M., Davies, R. L.,
        Kuntschner, H. \& Colless, M. 2006, MNRAS, 370, 1213


\bibitem[Chiboucas(2010)]{chi10} Chiboucas, K. et al.\ 2010, ApJ, 723, 251

\bibitem[Chilingarian(2008)]{chl10} Chilingarian I. V. \& Mamon, G. A. 2008, MNRAS, 385, L83

\bibitem[Da Rocha(2011)]{daroch11} Da Rocha, C., Mieske, S., Georgiev, I. Y., Hilker, M., 
        Ziegler, B. L., \& Mendes de Oliviera, C. 2011, A\&A, 525, A86
        
\bibitem[Dariush(2007)]{dariuch07} Dariush, A., Khosroshahi, H. G., Ponman, T. J., Pearce,
        F., Raychaudhury, S., \& Hartley, W. 2007, MNRAS, 382, 433 
        
        
\bibitem[Drinkwater(2000)]{dri00} Drinkwater, M. J. et al.\ 2000, PASA, 17, 227        

\bibitem[Evstigneeva et al.(2007)]{evs07} Evstigneeva, E. A., Drinkwater, M. J., 
Jurek, R., Firth, P., Jones, J. B., Gregg, M. D., \& S. Phillipps, S. 2007, MNRAS, 378, 1036

\bibitem[Evstigneeva et al.(2007)]{evs07a} Evstigneeva, E. A., Gregg, M. D., Drinkwater, M. J., 
\& Hilker, M 2007, AJ, 133, 1722

\bibitem[Evstigneevaetal(2008)]{evs08} Evstigneeva, E. A. et al.\ 2008, AJ, 136, 461

\bibitem[Harris et al.(2006)]{Har06} Harris, W. E., Whitmore, B. C., Karakla, D., Okon, W., 
        Baum, W. A., Hanes, D. A., \& Kavelaars, J. J. 2006, ApJ, 636, 90

\bibitem[Harris(2009)]{Har09} Harris, W. E. 2009, ApJ, 699, 254 

\bibitem[Hasegan et al.(2005)]{Has05} Hasegan, M., et al. 2005, ApJ, 627, 203

\bibitem[Hau(2009)]{Hau09} Hau, G. K., et al.\ 2009, MNRAS, 394, L97 

\bibitem[Hilker(1999)]{hil99}Hilker, M., et al.\ 1999, A\&AS, 134, 75

\bibitem[King(1962)]{kin62}King, I. R. 1962, AJ, 67, 471

\bibitem[King(1966)]{kin66}King, I. R. 1966, AJ, 71, 64

\bibitem[Laidler et al(2005)] {laid05} Laidler, V. et al. 2005, 
Synphot User's Guide, Version 5.0 (Baltimore, MD: STScI)

\bibitem[Larsen(1999)] {lar99} Larsen, S. S. 1999, A\&AS, 139, 393

\bibitem[Lintott et al.(2011)] {lin11} Lintott, C. et al.\ 2011, MNRAS, 410, 166

\bibitem[Mack(2003)]{mac03} Mack, J. et al.\ 2003, ACS Data Handbook,
Version 2.0, Baltimore, STScI

\bibitem[McLaughlin(2000)]{mcl00} McLaughlin, D. E. 2000, ApJ, 539, 618

\bibitem[Madrid et al. (2010)]{mad10} Madrid, J. P., et al. 2010, ApJ, 722, 1707 (Paper I)

\bibitem[Mieske(2004)]{mie04} Mieske, S., et al.\ 2004, AJ, 128, 1529

\bibitem[Mieske(2006)]{mie06} Mieske, S., et al.\ 2006, AJ, 131, 2442

\bibitem[Mieske(2007)]{mie07} Mieske, S., et al.\ 2007, A\&A, 472, 111

\bibitem[Misgeld(2008)]{mis08} Misgeld, I., Mieske, S., \& Hilker, M.\ 2008, A\&A, 486, 697

\bibitem[Mulchaey \& Zabludoff(1999)]{mul99} Mulchaey, J. S. \& Zabludoff, A. I. 1999, ApJ, 514, 133

\bibitem[Norris \& Kannappan(2011)]{nor11} Norris, M. A. \& Kannappan, S. J. 2011, astro-ph 1102.0001

\bibitem[Peng et al(2002)]{peng02} 	Peng, C. Y., Ho, L. C., Impey, C. D., Rix, H-W. 2002, AJ, 124, 266

\bibitem[Peng et al(2006)]{peng06} 	Peng, E. W. et al. 2006, ApJ, 639, 95

\bibitem[Peng et al(2009)]{peng09} 	Peng, E. W. et al. 2009, ApJ, 703, 42

\bibitem[Price(2009)]{pri09} Price, J. et al. 2009, MNRAS, 397, 1816

\bibitem[Romanowsky(2009)]{rom09} Romanowsky, A. J. et al.\ 2009, AJ, 137, 4956

\bibitem[Sales et al(2007)]{sal07} Sales, L. V., Navarro, J. F., Lambas, D. G., White, S. D. M., Croton, D. J. 2007, MNRAS, 382, 1901

\bibitem[Sersic(1968)]{ser68} S\'ersic, J.L. 1968, Atlas de Galaxias Australes (C\'ordoba: Obs. Astron\'om.)

\bibitem[Sirianni(2005)]{sir05} 
        Sirianni, M., et al. 2005, PASP, 117, 1049
        
\bibitem[Skrutskie(2006)]{skr06} 
        Skrutskie, M., F., et al. 2006, AJ, 131, 1163        
               
        
\bibitem[Springel et al.(2005)]{spr05} 
        Springel, V., et al. 2005, Nature, 435, 629    
        
\bibitem[Strader et al.(2006)]{str06} 
        Strader, J., Brodie, J. P., Spitler, L., \& Beasley, M. A. 2006, AJ, 132, 1593   
        
\bibitem[Wehner(2007)]{weh07} Wehner, E. M. H. \& Harris, W. E. 2007, ApJ, 668, L35        

\bibitem[Wright(2006)]{wri06}  
        Wright, E. L. 2006, PASP, 118, 1711
        

\end{thebibliography}
\end{document}